\documentclass[a4paper,12pt]{article}
\usepackage{amssymb,amsmath}

\begin{document}

\title{Non-singular $G_{2}$ stiff fluid cosmologies}

\author{L. Fern\'andez-Jambrina \\
E.T.S.I. Navales\\ Universidad Polit\'ecnica de Madrid \\ 
Arco de la Victoria s/n \\ E-28040 Madrid, Spain \\
and\\
L.M. Gonz\'alez-Romero\\
Departamento de F\'\i sica Te\'orica II\\ 
Facultad de Ciencias F\'\i sicas \\
Universidad Complutense de Madrid \\ 
Avenida Complutense s/n \\ E-28040 Madrid, Spain}

\date{\today}
\maketitle

 \begin{abstract}
     In this paper we analyse Abelian diagonal orthogonally transitive 
     spacetimes with spacelike orbits for which the matter content is a stiff 
     perfect fluid. The Einstein equations are cast in a suitable form for
     determining their geodesic completeness. A
     sufficient condition on the metric of these spacetimes 
     is obtained, that is fairly easy to check and to implement in 
     exact solutions. These results confirm that non-singular 
     spacetimes are abundant among stiff fluid cosmologies.
 \end{abstract}

 \noindent PACS {04.20.Dw, 04.20.Ex, 04.20.Jb}

\section{Introduction}

After the discovery of the first non-singular perfect fluid cosmological model by 
Senovilla \cite{seno}, the possibility of constructing regular 
cosmologies was renewed. The interest 
for regular cosmologies had stifled for nearly thirty years due to
the powerful singularity theorems (cfr. for instance \cite{HE}, 
\cite{beem}), which seemed to preclude such spacetimes under very 
general requirements, such as chronology protecting, energy and 
generic conditions. The open way to regular cosmologies was found in 
the violation of some technical premises of the theorems. For 
instance, in \cite{chinea} it was shown that the Senovilla spacetime 
did not possess a compact achronal set without edge and could not 
have closed trapped surfaces.

However, the first results were not encouraging. The extension of the 
Senovilla solution to a family of spacetimes left the set of regular 
models limited to a zero-measure subset surrounded 
by spacetimes with Ricci and Weyl curvature singularities \cite{ruiz}. During the 
following decade only a few new non-singular cosmologies were added 
to the list \cite{grg}.

Another estrategy to approach singularities arose with the publication 
of regularity theorems \cite{miguel}, \cite{manolo}, \cite{nondiag}. 
Whereas singularity theorems stated general sufficient conditions for the 
appearance of singularities, these theorems aimed the contrary, namely 
particular conditions to achieve regular spacetimes.

The application of the conclusions of \cite{manolo} to a restricted family of stiff fluids 
provided an unexpected result. The set of known non-singular 
perfect fluid cosmologies was enlarged with a huge family depending 
on two nearly arbitrary functions \cite{wide}. 

The purpose of this paper is the 
extension of those results to determine which spacetimes among 
Abelian diagonal orthogonally transitive spacetimes with spacelike 
orbits and with a stiff 
fluid as matter content are non-singular. Instead of restricting to 
an integrable family of solutions of the Einstein equations, we analyse 
the whole set of diagonal cylindrical stiff fluid spacetimes with a 
spacelike transitivity surface element.

With this aim in mind we write in Section \ref{equations} the Einstein equations for such 
spacetimes and we cast them in a form suitable for the application of 
the theorems. The analysis of the restrictions imposed by regularity 
conditions is done in Section \ref{theorems}. Finally in Section 
\ref{conclusions} we check the possibility of constructing regular 
spacetimes with non-vaninishing matter scalar space averages on Cauchy 
hypersurfaces in order to support the validity of a regularity 
conjecture by Senovilla \cite{senoray}.

\section{Equations for $G_{2}$ stiff fluid spacetimes\label{equations}}

As it has been stated in the introduction, we shall focus on 
spacetimes endowed with an Abelian orthogonally transitive group of 
isometries $G_{2}$ acting on spacelike surfaces, since this is the 
framework where most non-singular spacetimes have been found so far. We 
 further impose that generators for the group can be found 
that are mutually orthogonal. We  follow \cite{manolo} for 
writing the Einstein equations for such spacetimes using a formalism 
based on differential forms.

If the generators of the isometry group are chosen to be $\{\xi, 
\eta\}$, we may write an orthonormal tetrad,  
$\{\theta^{0}, \theta^{1}, \theta^{2} ,\theta^{3} \}$, where just 
$\theta^{2}$ and $\theta^{3}$ lie in ${\rm lin} \{\xi, \eta\}$. We may 
impose that these 1-forms be Lie-invariant under the isometry group 
\cite{lie}. The metric is written as 

\begin{equation}ds^2 = -\theta^0 \otimes \theta^0 + \theta^1 \otimes \theta^1 +
\theta^2 \otimes \theta^2 + \theta^3 \otimes \theta^3.\end{equation}

Making use of the spacelike congruence for $\theta^2$ and its 
kinematical quantities, we may define the tetrad basis according to 
the vanishing torsion equations, 

\begin{subequations}
    \begin{eqnarray}
       d \theta^0 & = &  \nu \wedge \theta^1
       \label{vt1}  \\
       d \theta^1 & = & \nu \wedge \theta^0
       \label{vt2}  \\
       d \theta^2 & = &  \alpha \wedge \theta^2 
       \label{vt3}  \\
       d \theta^3 & = & (\beta - \alpha) \wedge \theta^3 
       \label{vt4},
\end{eqnarray}
\end{subequations}
where $\nu$ is just a connection in the $\theta^0-\theta^1$ 
subspace, $\alpha$ is an ``acceleration'' for $\theta^2$ and $\beta$ 
is related to the expansion of the surface element in the 
$\theta^2-\theta^3$ subspace, since $d(\theta^2\wedge 
\theta^3)=\beta\wedge \theta^2\wedge \theta^3$. 

The integrability conditions for these equations are easily obtained 
by exterior differentiation of the system,

\begin{subequations}
\begin{eqnarray}
    d \beta &=& 0,    
    \label{1b1}  \\
    d \alpha &=& 0.    
    \label{1b2} 
\end{eqnarray}
\end{subequations}

Finally, Einstein field equations are written in terms of these 
differential forms as an exterior system,

\begin{subequations}
\begin{eqnarray}
       && d \ast \alpha + \beta \wedge \ast \alpha 
        =  \left(\frac{1}{2}T-T_{22}\right)\theta^0\wedge\theta^1, 
       \label{fe2}  \\
       && d \ast \beta + \beta \wedge \ast \beta  
        =   \left(T_{11}-T_{00}\right)\theta^0\wedge\theta^1,
       \label{fe3}  \\
       && d \nu + \alpha \wedge \ast \alpha -  \beta 
       \wedge \ast \alpha  
       = \frac{1}{2}\left(T_{00}-T_{11}+T_{22}+T_{33}\right)\theta^0\wedge\theta^1,
       \label{fe4}  \\
       && d \ast \tilde{\beta} + \beta \wedge \ast \tilde{\beta} + 
       2 (\alpha - \beta) \wedge \ast \tilde{\alpha} + 2 \nu \wedge \tilde{\beta}
	 =  \left(T_{00}+T_{11}\right)\theta^0\wedge\theta^1,
       \label{fe5}  \\
       && d  \tilde{\beta} + \beta \wedge \tilde{\beta} +  
       2 (\alpha - \beta)\wedge  \tilde{\alpha} + 2 \nu \wedge \ast 
	\tilde{\beta} 
	=  2 T_{01}\theta^0\wedge\theta^1,
       \label{fe6}
\end{eqnarray}
\end{subequations}
for a matter content defined by the energy-momentum tensor 
$T=T_{ab}\theta^a\otimes \theta^b$. 

The tilde denotes a reflection in the $\theta^0-\theta^1$ 
subspace, that is, if $\alpha=a\theta^0+b\theta^1$, then 
$\tilde \alpha=a\theta^0-b\theta^1$. The $\ast$ denotes the Hodge 
duality operator in the same subspace, $\ast\alpha=-a\theta^1-b\theta^0$.

Integration of the first Bianchi equations (\ref{1b1},\ref{1b2}), 

\begin{subequations}
\begin{eqnarray}
    \alpha & = & -dU,\\
    \beta & = & \frac{d\ln\rho}{\rho},
\end{eqnarray}
\end{subequations}

allows integration of Cartan equations in terms of two functions, 
$z$, $\phi$,
\begin{equation}
    \theta^2=e^{-U}\,dz,\quad \theta^3=\rho e^U\,d\phi,
    \end{equation}
that we take as coordinates in order to write the metric in a 
conventional form,
\begin{equation}
g=e^{2K}(-dt^2+dr^2)+e^{-2U}dz^2+\rho^2e^{2U}d\phi^2.\label{metric}
\end{equation}

The coordinates are adapted to the Killing fields, so that 
$\xi=\partial_{z}$, $\eta=\partial_{\phi}$. The non-ignorable 
coordinates $t$, $r$ are chosen so that the metric is isotropic in 
the subspace spanned by $\theta^0$ and $\theta^1$,
\begin{equation}
    \theta^0=e^{K}dt,\quad \theta^1=e^{K}dr.
    \end{equation}
    
The range for these coordinates is 
the usual one,
\begin{equation}
    -\infty<t,z<\infty,\ 0<r<\infty,\ 0<\phi<2\pi,
\end{equation}
if we require the spacetime to be cylindrically symmetric. The 
remaining metric functions, 
$K$, $U$ and $\rho$, depend just on $t$ and $r$. 

The connection in this case is $\nu=*dK$.

This is the general framework for an orthogonally transitive diagonal 
spacetime with spacelike 
orbits. If the matter content is a perfect fluid with 4-velocity 
$u$, pressure $p$ and density $\mu$, the Bianchi equations for such 
energy-momentum tensor,

\begin{equation}
    T=\mu u\otimes u+p\,(g+u\otimes u),
\end{equation} 
may be written in compact expressions involving the kynematical 
1-forms,

\begin{subequations}
\begin{eqnarray}
       && d u + \frac{1}{\mu + p} d p \wedge u  =  0,
       \label{bi1}  \\
       && d \ast u + \left(\beta + \frac{d \mu}{\mu + p}  \right) \wedge \ast u  = 
       0, 
       \label{bi2}
\end{eqnarray}
\end{subequations}
which state that the fluid is irrotational.

We might choose $\theta^0=u$ for writing the Einstein equations, as it 
was done in \cite{wide}, but since we aim full generality, 
 we shall not follow that way and explore arbitrary possibilities of alignment 
for this 1-form. 
Preserving the unitarity of $u$, we may parametrize it in terms of a 
function $\xi$,

\begin{equation}
    u=-\theta^0\,\cosh \xi -\theta^1\,\sinh\xi,
\end{equation} 
so that the Einstein equations for a perfect fluid take the following 
form

\begin{subequations}
\begin{eqnarray}
&& 
U_{tt}-U_{rr}+\frac{1}{\rho}(U_{t}\rho_{t}-U_{r}\rho_{r})=\frac{p-\mu}{2}e^{2K},\label{U}
\\
&& \rho_{tt}-\rho_{rr}=(\mu-p)\rho e^{2K},\label{rho}
\\
&& 
K_{t}\rho_{r}+K_{r}\rho_{t}=\rho_{tr}+U_{t}\rho_{r}+U_{r}\rho_{t}+2 
\rho U_{t}U_{r}\nonumber\\&&+e^{2K}\rho \frac{\mu+p}{2}\sinh2\xi,\label{Kt}
\\
&& 
K_{t}\rho_{t}+K_{r}\rho_{r}=\frac{\rho_{tt}+\rho_{rr}}{2}+
U_{t}\rho_{t}+U_{r}\rho_{r}\nonumber\\&&
+\rho\left(U_{t}^2+U_{r}^2+ 
e^{2K}\frac{\mu+p}{2}\cosh2\xi\right),\label{Kr}
\\
&& 
K_{rr}-K_{tt}+\frac{U_{r}\rho_{r}-U_{t}\rho_{t}}{\rho}+U_{r}^2-U_{t}^2=\frac{\mu+
p}{2}\,e^{2K}, \label{nu}
\end{eqnarray}
\end{subequations}
and the energy-momentum conservation laws yield

\begin{subequations}
\begin{eqnarray}
&& 
K_{r}-\xi_{t}+\frac{p_{r}\cosh^2\xi+
(\mu_{t}-p_{t})\sinh\xi\cosh\xi-\mu_{r}\sinh^2\xi}{\mu+p}\nonumber\\
&&+\frac{\rho_{t}\cosh\xi-\rho_{r}\sinh\xi}{\rho}
\sinh\xi=0,\label{p}
\\
&& K_{t}-\xi_{r}+\frac{\mu_{t}\cosh^2\xi+
(p_{r}-\mu_{r})\sinh\xi\cosh\xi-p_{t}\sinh^2\xi}{\mu+p}\nonumber\\
&&+\frac{\rho_{t}\cosh\xi-\rho_{r}\sinh\xi}{\rho}
\cosh\xi=0.\label{mu}
\end{eqnarray}
\end{subequations}

The system of equations becomes much simpler if we restrict  to 
stiff fluids, $\mu=p$, 

\begin{subequations}
\begin{eqnarray}
    && U_{tt}-U_{rr}+\frac{1}{\rho}(U_{t}\rho_{t}-U_{r}\rho_{r})=0,\label{U1}
    \\
    && \rho_{tt}-\rho_{rr}=0,\label{rho1}
    \\
    && 
    \frac{K_{t}\rho_{r}+K_{r}\rho_{t}}{\rho}=\frac{\rho_{tr}+U_{t}\rho_{r}+U_{r}\rho_{t}}{\rho}+2 
    U_{t}U_{r}+e^{2K} p\sinh2\xi,\label{Kt1}
    \\
    && 
    \frac{K_{t}\rho_{t}+K_{r}\rho_{r}}{\rho}=\frac{\rho_{tt}+\rho_{rr}}{2\rho}+
    \frac{U_{t}\rho_{t}+U_{r}\rho_{r}}{\rho}
    +U_{t}^2+U_{r}^2\nonumber\\&&+ e^{2K}p\cosh2\xi,\label{Kr1}
    \\
    && 
    K_{rr}-K_{tt}+\frac{U_{r}\rho_{r}-U_{t}\rho_{t}}{\rho}+U_{r}^2-U_{t}^2=
    p\,e^{2K}, \label{nu1}\\
    && 
    K_{r}-\xi_{t}+\frac{p_{r}}{2p}+\frac{\rho_{t}\cosh\xi-\rho_{r}\sinh\xi}{\rho}
    \sinh\xi=0,\label{p1}
    \\
    && K_{t}-\xi_{r}+\frac{p_{t}}{2p}
    +\frac{\rho_{t}\cosh\xi-\rho_{r}\sinh\xi}{\rho}
    \cosh\xi=0.\label{mu1}
\end{eqnarray}
\end{subequations}

The reason why the stiff fluid equations are easy to integrate is that 
the metric functions $U$, $\rho$ decouple from the pressure, which 
only appears in the equations for the conformal factor $K$. Therefore 
the stiff fluid case is fairly similar to vaccuum and can be 
generated from this one. 

A further simplification can be obtained if we take $\rho$ as 
coordinate. This is fully compatible with an isotropic 
parametrization, since equation (\ref{rho1}),

\begin{eqnarray*}
    0=d*\beta+\beta\wedge *\beta=
    d\left(\frac{*d\rho}{\rho}\right)+\frac{d\rho\wedge*d\rho}{\rho^2}=\frac{d*d\rho}{\rho},
    \end{eqnarray*}
states that $*d\rho$ is also an exact differential form.

We take $\rho=r$ as an spatial coordinate, since every known 
non-singular solution has a surface element with spacelike gradient. 
With this choice of coordinates the differential system becomes even 
simpler,

\begin{subequations}
\begin{eqnarray}
&& U_{tt}-U_{rr}-\frac{U_{r}}{r}=0,\label{U2}
\\ && 
K_{t}=U_{t}+2r U_{t}U_{r}+e^{2K}pr\sinh2\xi,\label{Kt2}
\\ && 
K_{r}=U_{r}+r(U_{t}^2+U_{r}^2)+e^{2K}pr\cosh2\xi,\label{Kr2}
\\ && 
K_{rr}-K_{tt}+\frac{U_{r}}{r}+U_{r}^2-U_{t}^2=pe^{2K}, \label{nu2}
\\
&& 
K_{r}-\xi_{t}+\frac{p_{r}}{2p}-\frac{\sinh^2\xi}{r}=0,\label{p2}
\\
&& K_{t}-\xi_{r}+\frac{p_{t}}{2p}-\frac{\sinh\xi\cosh\xi}{r}=0.
\label{mu2}
\end{eqnarray}
\end{subequations}

The integrability condition, $K_{rt}=K_{tr}$, for (\ref{Kt2}, 
\ref{Kr2}) requires that a combination of functions be an exact 
differential form,

\begin{equation}\label{exact}
    dH= e^{2K}rp(\sinh 2\xi\,dt+\cosh 2\xi\,dr),
\end{equation}
from which we can read $\xi$ and the pressure, if $K$ is known,

\begin{equation}\label{pdef}
    \tanh 2\xi= \frac{H_{t}}{H_{r}},\qquad 
    |p|=\frac{e^{-2K}}{r}\sqrt{H_{r}^2-H_{t}^2}.
\end{equation}

The integrability of $dH$ is also a consequence of the 
energy-momentum conservation equations (\ref{p2}, \ref{mu2}).

For consistency these expressions imply that the gradient of $H$ be 
spacelike and that $H_{r}$ be positive in order to have positive 
pressure. 

The simple case, $\xi=0$, for which $u$ is parallel to the time 
direction corresponds to $H=\gamma r^2/2$, where $\gamma$ is a 
positive constant.

The remaining system of differential equations,

\begin{subequations}
\begin{eqnarray}
    && U_{tt}-U_{rr}-\frac{U_{r}}{r}=0,\label{U3}
    \\ && 
    H_{rr}-H_{tt}=\frac{\sqrt{H_{r}^2-H_{t}^2}}{r}, \label{H}
    \\ && 
    K_{t}=U_{t}+2r U_{t}U_{r}+H_{t},\label{Kt3}
    \\ && 
    K_{r}=U_{r}+r(U_{t}^2+U_{r}^2)+H_{r},\label{Kr3}    
\end{eqnarray}
\end{subequations}
is formed by a reduced wave equation in polar coordinates for $U$ on the plane 
and a non-linear wave equation for $H$. Once these equations are 
solved, we are left with a quadrature for $K$. The integrability of 
this quadrature is guaranteed by the other equations.

As it has already been stated, these equations are pretty similar to 
those of vaccuum. The only diference is the additional conformal 
factor defined by $H$.

Regularity of the metric at the axis $r=0$ is already implicit in the 
equations, provided that metric functions are regular. Following 
\cite{kramer}, we have a regular axis whenever
\begin{equation}
    \lim_{r\to0}\frac{\langle 
	\textrm{grad}\,\Delta,\textrm{grad}\,\Delta\rangle}{4\Delta}=
	e^{2(U-K)}|_{r=0}=1, 
	\quad 
	\Delta=\langle\partial_{\phi},\partial_{\phi}\rangle=r^2e^{2U}.
    \label{axis}
\end{equation}

But according to equations (\ref{exact}, \ref{Kt3}, \ref{Kr3}), $K$ and $U$ are 
equal at the axis, except for a constant, since 

\begin{equation}\label{Kaxis}
    K_{r}(t,0)=U_{r}(t,0),\qquad K_{t}(t,0)=U_{t}(t,0),
\end{equation} if pressure and $K$ are regular functions, so that 
$H_{t}(t,0)=0=H_{r}(t,0)$, and therefore condition (\ref{axis}) is fulfilled 
either by taking the constant equal to zero or conveniently 
rescaling the angular coordinate.

The problem of obtaining solutions for $H$ is solved by the 
Wainright-Ince-Marshman formalism \cite{Wain}. Solutions to (\ref{H}) 
may be generated from solutions of the reduced wave equation on the 
plane with timelike gradient,

\begin{equation}
    \sigma_{tt}-\sigma_{rr}-\frac{\sigma_{r}}{r}=0,\quad 
    \sigma_{t}^2-\sigma_{r}^2>0,
    \end{equation}
by a quadrature identical to the one which defines $K-U$ in the vacuum 
case,

\begin{subequations}
\begin{eqnarray}
 && H_{t}=2r \sigma_{t}\sigma_{r},\label{Ht}
    \\ && 
    H_{r}=r(\sigma_{t}^2+\sigma_{r}^2).\label{Hr}    
\end{eqnarray}
\end{subequations}

The functions $H$ generated by this mechanism have trivially a 
spacelike gradient and positive radial derivative. The fluid 
properties may be read directly from the generating function,

\begin{equation}\label{pwain}
    \tanh 2\xi= 
    \frac{2\sigma_{t}\sigma_{r}}{\sigma_{t}^2+\sigma_{r}^2},\qquad 
    p=e^{-2K}(\sigma_{t}^2-\sigma_{r}^2).
\end{equation}

The function that generates the $\xi=0$ case is 
$\sigma=\sqrt{\gamma}t$.

Using this formalism, the remaining system of equations is formed by 
a quadrature and two reduced wave equations,

\begin{subequations}
\begin{eqnarray}
    && U_{tt}-U_{rr}-\frac{U_{r}}{r}=0,\label{U4}
    \\ && 
    \sigma_{tt}-\sigma_{rr}-\frac{\sigma_{r}}{r}=0, \label{s4}
    \\ && 
    K_{t}=U_{t}+2r U_{t}U_{r}+2r \sigma_{t}\sigma_{r},\label{Kt4}
    \\ && 
    K_{r}=U_{r}+r(U_{t}^2+U_{r}^2)+r(\sigma_{t}^2+\sigma_{r}^2).\label{Kr4}    
\end{eqnarray}
\end{subequations}

\section{Non-singular models\label{theorems}}

We have obtained a fairly simple system of equations 
(\ref{U4}-\ref{Kt4}) that will be useful for analysing the regularity 
of the solutions. Following \cite{HE} we take causal geodesic 
completeness as our definition for regularity. 

Even if we have regular metric components, geodesic 
completeness of the spacetime is not guaranteed and we have to check 
explicitly that every timelike and lightlike geodesic in the 
spacetime can be extended to all values of the affine parameter, that 
is, in the parametrization for which the geodesic equations take the 
form

\begin{equation}
    \ddot x^i +\Gamma^i_{jk}\dot x^j\dot x^k,
\end{equation}
in terms of the Christoffel symbols.

Fortunately, results concerning causal geodesic completeness of diagonal 
Abelian orthogonally 
transitive spacetimes  have already been obtained in \cite{manolo}. 
The conclusions of that paper may be condensed in two theorems. We 
follow the simplified version of \cite{wide}.

\begin{description}
\item[Theorem 1:] A diagonal Abelian orthogonally transitive spacetime 
with spacelike orbits endowed with a metric in the form (\ref{metric}) with $C^2$
metric functions  $K,U,\rho$, where $\rho$ has a spacelike gradient, is future causally geodesically complete 
provided that along causal geodesics:
\begin{enumerate}
\item For large values of $t$ and increasing $r$, 
\begin{enumerate}
    \item $(K-U-\ln\rho)_{r}+(K-U-\ln\rho)_{t}\ge 0$, and either 
    $(K-U-\ln\rho)_{r}\ge 0$  or $|(K-U-\ln\rho)_{r}|\lesssim 
    (K-U-\ln\rho)_{r}+(K-U-\ln\rho)_{t}$.
    \item $K_{r}+K_{t}\ge 0$, and either $K_{r}\ge 0$ or 
    $|K_{r}|\lesssim K_{r}+K_{t}$.
    \item $(K+U)_{r}+(K+U)_{t}\ge 0$, and either $(K+U)_{r}\ge 0$ 
    or $|(K+U)_{r}|\lesssim (K+U)_{r}+(K+U)_{t}$.
\end{enumerate}

\item \label{tt} For large values of  $t$, a constant $b$ exists such that 
\\
$\left.\begin{array}{c}K(t,r)-U(t,r)\\2\,K(t,r)\\K(t,r)+U(t,r)+\ln\rho(t,r)
\end{array}\right\}\ge-\ln|t|+b.$

\end{enumerate}\end{description}

\begin{description}
\item[Theorem 2:] A diagonal Abelian orthogonally transitive spacetime 
with spacelike orbits endowed with a metric in the form (\ref{metric}) with $C^2$
metric functions  $K,U,\rho$, where $\rho$ has a spacelike gradient, is past causally geodesically complete 
provided that along causal geodesics:
\begin{enumerate}
\item For small values of $t$ and increasing $r$, 
\begin{enumerate}
    \item $(K-U-\ln\rho)_{r}-(K-U-\ln\rho)_{t}\ge 0$, and either 
    $(K-U-\ln\rho)_{r}\ge 0$  or $|(K-U-\ln\rho)_{r}|\lesssim 
    (K-U-\ln\rho)_{r}-(K-U-\ln\rho)_{t}$.
    \item $K_{r}-K_{t}\ge 0$, and either $K_{r}\ge 0$ or 
    $|K_{r}|\lesssim K_{r}-K_{t}$.
    \item $(K+U)_{r}-(K+U)_{t}\ge 0$, and either $(K+U)_{r}\ge 0$ 
    or $|(K+U)_{r}|\lesssim (K+U)_{r}-(K+U)_{t}$.
\end{enumerate}

\item \label{tt2} For small values of  $t$, a constant $b$ exists such that 
\\
$\left.\begin{array}{c}K(t,r)-U(t,r)\\2\,K(t,r)\\K(t,r)+U(t,r)+\ln\rho(t,r)
\end{array}\right\}\ge-\ln|t|+b.$

\end{enumerate}\end{description}

Therefore now we just have to verify under which conditions these 
theorems can be applied to stiff fluid spacetimes. Since the theorems 
do not make use of Einstein equations, it is expected that when we 
take them into account the conditions will not be so restrictive as they seem.

We begin with future-pointing geodesics. The first part of the theorem is a set of conditions on the 
derivatives of the metric functions:

\begin{enumerate}
    \item  
    \begin{enumerate}
        \item \label{condef} From (\ref{Kt3},\ref{Kr3}) we obtain 
	\begin{displaymath}
            (K-U-\ln\rho)_{t}+(K-U-\ln\rho)_{r}=
	    r (U_{t}+U_{r})^2+H_{t}+H_{r}-\frac{1}{r}.
        \end{displaymath}
	The sum of the derivatives of $H$ is always positive, since $H_{r}>|H_{t}|$ in 
	order to have positive pressure. In fact, this is the $r 
	(\sigma_{t}+\sigma_{r})^2$ term in the Wainright-Ince-Marshman formalism. 
	
	This expression is positive if either $|U_{t}+U_{r}|$ or $H_{t}+H_{r}$ 
	($|\sigma_{t}+\sigma_{r}|$ in the Wainright-Ince-Marshman formalism) 
	does not decrease as $1/r$ or faster for large values of $t$ and $r$. That is, we need 
	either $U$ or $H$ to overcome the negative term.
	
	Under such conditions, the second part of the premise,
	
	\begin{displaymath}
            (K-U-\ln\rho)_{r}=
	    r (U_{t}^2+U_{r}^2)+H_{r}-\frac{1}{r}\ge0,
        \end{displaymath} is also satisfied.
	
        \item   Once (\ref{condef}) is fulfilled, this condition, 
        \begin{displaymath}
            K_{t}+K_{r}=U_{t}+U_{r}+r (U_{t}+U_{r})^2+H_{t}+H_{r}>0,
        \end{displaymath} 
	is trivial, since the only possible negative contribution would be 
	that of $U_{t}+U_{r}$ and this is counteracted by $H_{t}+H_{r}$ if 
	it decreases as $1/r$ or faster, or by $r(U_{t}+U_{r})^2$ if it does 
	not.
	
	Following a similar line of thought we also conclude that 
	\begin{displaymath}
            K_{r}=U_{r}+r (U_{t}^2+U_{r}^2)+H_{r}.
        \end{displaymath} is positive for large values of $t$ and $r$.
    
        \item The last set of conditions on the derivatives, 
        \begin{displaymath}
            (K+U)_{t}+(K+U)_{r}=2(U_{t}+U_{r})+r 
            (U_{t}+U_{r})^2+H_{t}+H_{r}\ge0,
        \end{displaymath}
	\begin{displaymath}
	    K_{r}+U_{r}=2U_{r}+r U_{r}^2+H_{r}\ge0,
	\end{displaymath} 
	is also a consequence of (\ref{condef}). 
	
    \end{enumerate}
    
    Therefore the first part of the theorem is satisfied if 

    \begin{equation}
	    \left.
	    \begin{array}{c}
		r^{1-\varepsilon}|U_{r}+U_{t}|  \\
		\textrm{or}\\
		r^{1-\varepsilon}(H_{r}+H_{t})
	    \end{array}\right\}\not\to 0
	    \label{cond1}
	\end{equation}    for large values of $t$ and $r$.
    
    The conclusion for past-pointing geodesics is quite similar. We 
    just have to change the sign of the time derivatives.
    
    \begin{equation}
	\left.
	\begin{array}{c}
	    r^{1-\varepsilon}|U_{r}-U_{t}|  \\
	    \textrm{or}\\
	    r^{1-\varepsilon}(H_{r}-H_{t})
	\end{array}\right\}\not\to 0
	\label{cond2}
    \end{equation}
for large values of $r$ and small values of $t$.    
    
For instance, these restrictions are trivial for the $\xi=0$ case, 
since $H=\gamma r^2/2$ does not decrease.

\item The dependence on the matter content of the 
spacetime may be removed from these conditions, since we may write
    \begin{eqnarray}\label{KU}
        K(t,r)&=&
	U(t,r)+\int_{0}^rdr'\,\big(r' 
	U_{r}^2(t,r')+r'U_{t}^2(t,r')+H_{r}(t,r')\big)
	\nonumber\\ &=&U(t,0)+\int_{0}^rdr'\,K_{r}(t,r'),
    \end{eqnarray}
and according to (\ref{Kr3}) or (\ref{Kr4}) $K_{r}$ is a positive term if 
the first part of the theorem is satisfied.

    \begin{enumerate}
        \item  The first condition is tautological since 
	\begin{displaymath}
	     K(t,r)-U(t,r)=
	    \int_{0}^{r}dr'\,\big(r' 
	U_{r}^2(t,r')+r'U_{t}^2(t,r')+H_{r}(t,r')\big)>0.
	\end{displaymath}
    
        \item  For geodesics along the axis, this condition requires 
        for large values of the time coordinate that
	\begin{equation}\label{ax}
	    K(t,0)=U(t,0)\ge -\frac{1}{2}\ln |t|+b,
	\end{equation}
	and for general geodesics the only difference is the positive term 
	in (\ref{KU}). Therefore (\ref{ax}) is the only restriction for all 
	geodesics.
	    
        \item  The same restriction is achieved likewise when applied 
        to the expression $K+U+\ln\rho$.
    \end{enumerate}
\end{enumerate}

Therefore we are left with just three regularity conditions on the metric of an 
Abelian diagonal orthogonally transitive spacetime  with spacelike 
orbits and with a stiff 
perfect fluid as matter content. We may summarize these results in two 
theorems:

\begin{description}
\item[Theorem 3:] A cylindrical spacetime 
with a stiff perfect fluid as matter content, 
endowed with a metric in the form (\ref{metric}) with $C^2$
metric functions  $K,U,\rho$ is future geodesically complete 
if the gradient of the surface element is spacelike and
\begin{enumerate}
\item For large values of $t$, $U(t,0)\ge -\displaystyle\frac{1}{2}\ln |t|+b$.

\item Either $r^{1-\varepsilon}|U_{r}+ U_{t}|$ or 
$r^{1-\varepsilon}(H_{r}+ H_{t})$ does not tend to zero for large 
values of $t$ and $r$.

\end{enumerate}\end{description}

\begin{description}
\item[Theorem 4:] A cylindrical spacetime 
with a stiff perfect fluid as matter content, 
endowed with a metric in the form (\ref{metric}) with $C^2$
metric functions  $K,U,\rho$ is past geodesically complete 
if the gradient of the surface element is spacelike and
\begin{enumerate}
\item For small values of $t$, $U(t,0)\ge -\displaystyle\frac{1}{2}\ln |t|+b$.

\item Either $r^{1-\varepsilon}|U_{r}- U_{t}|$ or 
$r^{1-\varepsilon}(H_{r}- H_{t})$ does not tend to zero for small
values of $t$ and large values of $r$.

\end{enumerate}\end{description}

For vaccuum spacetimes both theorems hold just dropping the 
conditions on the derivatives of $H$.

The restrictions imposed by both theorems in order to have a 
non-singular spacetime are rather simple to implement, since $U$ is 
just a solution of the wave equation and $H$ is related to another one. 
We may state that regularity conditions are quite weak for 
stiff fluids, since it is very easy to provide solutions that fulfill 
such requirements. For instance:

\begin{description}
    \item[Corolary:] A metric with arbitrary $H$ and 
a function $U$ which grows for large $|t|$ and for large $r$ makes the spacetime 
geodesically complete. 
\end{description}

It is not difficult to derive such functions. The solutions to the 
reduced wave equation in the plane can be written as 
solutions of the initial value problem,

\begin{eqnarray}
    && U_{tt}-U_{rr}-\frac{U_{r}}{r}=0,\nonumber\\
    && U(0,r)=f(r),\quad U_{t}(0,r)=g(r).
    \end{eqnarray}
    
The solution to this problem can be written in closed form 
\cite{john},

\begin{eqnarray}
U(x,y,t)&=&\frac{1}{2\pi}\int_{0}^{2\pi}d\phi\int_0^tdR 
R\frac{g(x+R\cos\phi,y+R\sin\phi)}{\sqrt{t^2-R^2}}\nonumber\\ &+&
\frac{1}{2\pi}\frac{\partial}{\partial 
t}\int_{0}^{2\pi}d\phi\int_0^tdR 
R\frac{f(x+R\cos\phi,y+R\sin\phi)}{\sqrt{t^2-R^2}},
\end{eqnarray} 
for initial data $U(x,y,0)=f(x,y)$, $U_{t}(x,y,0)=g(x,y)$, taking into 
account that $f$ and $g$ are to have circular symmetry.

If we split $U$ in $U_{f}$ and $U_{g}$, the terms depending respectively on the 
initial data for $U$ and its derivative, we notice that $U_{f}$ is 
even in the time coordinate whereas $U_{g}$ is odd. This implies that 
if $U_{f}$ satisfies the first condition in theorem 3 for 
future-pointing geodesics, it will fulfill it for past-pointing 
geodesics too. On the contrary, if $U_{g}$ satisfies it for 
future-pointing geodesics, it may not fulfill it for past-pointing 
ones, unless it behaves for large values of $|t|$ slower than a 
logarithm. Therefore one is to require either that $U_{f}$ dominates over 
$U_{g}$ for large $|t|$ or that both terms behave slower than a 
logarithm in order to have non-singular behaviour.

\section{Discussion\label{conclusions}}

In this paper we have derived sufficient conditions for an Abelian 
diagonal orthogonally transitive spacetime  with spacelike 
orbits and with a stiff perfect fluid 
as matter content to be geodesically complete. One of the metric functions appears to be determinant for the 
regularity of the spacetime. These conditions are 
easy to check and do not mean much restriction on these spacetimes. 
 
This means that non-singular spacetimes are not as scarce as it was 
thought, considering the reduced list of geodesically complete 
perfect fluid cosmologies in the literature. Further work is needed 
with more generic symmetries and matter contents in order to clarify 
the issue, since stiff perfect fluids are rather peculiar. They may 
be interpreted as a massless scalar field and they are the limiting 
case for which a barotropic perfect fluid with linear equation of 
state satisfies every energy condition. These spacetimes  also fulfill the  
generic condition and are causally stable. 

The latter assert is true 
since they possess a cosmic time, which is the coordinate $t$. This 
coordinate has a timelike gradient everywhere. Therefore \cite{HE}, 
these spacetimes satisfy weaker causality conditions. For instance, 
the chronology condition is true for them and no closed causal curves 
are possible.

As it was stated in the introduction, the existence of these 
non-singular spacetimes is possible because they do not possess 
causally trapped sets. They obviously do not contradict then the 
singularity theorems. They just fall out of their scope.

Another interesting point that is worthwhile mentioning is that the 
regularity theorems appear to encourage a growing $K$ for large 
values of $|t|$. This seems to support a conjecture that states that 
the spatial average value of the pressure in non-singular spacetimes is 
zero \cite{senoray}, since $p$ decreases with large $K$ according to (\ref{pwain}),

\begin{equation}
    p=e^{-2K}(\sigma_{t}^2-\sigma_{r}^2).
\end{equation}

In our regular spacetimes, constant $t$ sheets are Cauchy 
hypersurfaces and we may write the whole system of equations as an 
initial value problem for $U$, $K$ and $H$ for any constant $t$. 
Without breaking the generality of the result, we may focus on 
$t=0$. The initial value problem can be stated as

\begin{subequations}
\begin{eqnarray}
    && U_{tt}-U_{rr}-\frac{U_{r}}{r}=0,
    \\ &&
    \sigma_{tt}-\sigma_{rr}-\frac{\sigma_{r}}{r}=0,
   \\ &&
    K_{t}=U_{t}+2r \left(U_{t}U_{r}+ \sigma_{t}\sigma_{r}\right),
    \\ && 
     U(0,r)=f(r),\quad U_{t}(0,r)=g(r),
    \\ && 
    \sigma(0,r)=f_{\sigma}(r),\quad 
    \sigma_{t}(0,r)=g_{\sigma}(r),
    \\ &&
    K(0,r)=h(r),
\end{eqnarray}
\end{subequations}
and the remaining equation in the system, 

\begin{equation}
        K_{r}=U_{r}+r(U_{t}^2+U_{r}^2+\sigma_{t}^2+\sigma_{r}^2),     
\end{equation}
is used to complete the initial data,

\begin{eqnarray}
    h(r)&=& U(0,0)+\int_{0}^{r}dr'\,K_{r}(0,r')\nonumber\\&=&
    f(r)+\int_{0}^{r}dr'\, 
    r'\left\{g(r)^2+f'(r)^2+g_{\sigma}(r')^2+f_{\sigma}'(r')^2\right\}.
\end{eqnarray}

In order to know the pressure on the hypersurface $t=0$ we just have 
to prescribe the initial data,

\begin{equation}
    p(0,r)=e^{-2h(r)}\left\{g_{\sigma}(r)^2-f_{\sigma}'(r)^2\right\}.
\end{equation}

We show that it must necessarily vanish at infinity if the spacetime 
is causally geodesically complete.

If the term  $g_{\sigma}^2-f_{\sigma}'^2$in the pressure does not tend to zero at 
infinity, the $\sigma$ terms would contribute to $h$ as $r^2$ (if 
$g_{\sigma}^2+f_{\sigma}'^2$ tends to a constant) or greater. Unless a 
negative $f$ overcomes this quadratic term, we would have a pressure 
decreasing as a Gaussian exponential and the average on $t=0$ would be 
zero.

But $f$ cannot beat a quadratic term, because the $f'$ term in the 
integral would mean a positive $r^4$ contribution to $h$, and we would 
have again a negative exponential. That is, if 
$g_{\sigma}^2-f_{\sigma}'^2$ does not vanish at infinity, it grows 
much slower than the 
exponential term decreases and the pressure tends to zero.

The only possibility we have left then is a positive exponential. This 
means a negative $h$. If we want $f$ to overcome just the $f'$ term in 
$h$, we require $|f(r)|\le \ln r$ for large values of $r$, a very 
narrow strip. 

But we also need to keep under control the $\sigma$ terms in $h$. 
They remain bounded for large values of $r$ if 
$r^{2}(g_{\sigma}^2+f_{\sigma}'^2)$ tends to zero. This means that 
the $\sigma$ term in the pressure decreases faster than $r^{-2}$.  
Admitting that $h(r)$ might behave as $-\ln r$ for large $r$, the 
exponential in the pressure would be a $r^2$ term, that cannot 
compensate the $\sigma$ term.

Therefore, pressure tends to zero for large $r$ on constant time 
hypersurfaces, thereby supporting Senovilla's conjecture in 
\cite{senoray}.

\section*{Acknowledgments}
The present work has been supported by Direcci\'on General de
Ense\~nanza Superior Project PB98-0772. The authors wish to thank
 F.J. Chinea,  F. Navarro-L\'erida and  M.J. Pareja 
for valuable discussions.

\end{document}